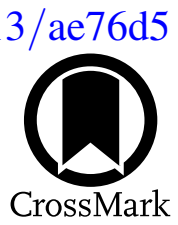

# Detection of $C_{60}$ Combination Bands in the Near-IR Spectrum of Tc 1

Morgan M. Giese[1,2], Vincent J. Esposito[3], Simon Van Schuylenbergh[1,2], Jan Cami[1,2], Els Peeters[1,2], Charmi Bhatt[1,2], Dries Van De Putte[1,2], A. G. G. M. Tielens[4], Michael J. Barlow[5], Jeronimo Bernard-Salas[6,7], Alessandra Candian[8], Bryan Changala[9,10], Nick L. J. Cox[6], Harriet L. Dinerstein[11], D. A. García-Hernández[12,13], Marco A. Gómez-Muñoz[14,15], Kay Justtanont[16], Kathleen E. Kraemer[17], Eric Lagadec[18], Arturo Manchado[12,13,19], Ana Monreal Ibero[20], Raghvendra Sahai[21], Ameek Sidhu[1,2], G. C. Sloan[22,23], N. C. Sterling[24], Jeremy R. Walsh[25], Roger Wesson[5,26], Joshua Cole Whitman[27], and Albert Zijlstra[28]

[1] Department of Physics & Astronomy, The University of Western Ontario, London, N6A 3K7, Canada; mgiese@uwo.ca
[2] Institute for Earth and Space Exploration, The University of Western Ontario, London, N6A 3K7, Canada
[3] Schmid College of Science and Technology, Chapman University, Orange, CA 92866, USA
[4] Astronomy Department, University of Maryland, College Park, MD 20742, USA
[5] Department of Physics and Astronomy, University College London, Gower St., London WC1E 6BT, UK
[6] ACRI-ST, Centre d'Etudes et de Recherche de Grasse (CERGA), 10 Av. Nicolas Copernic, 06130 Grasse, France
[7] INCLASS Common Laboratory, 10 Av. Nicolas Copernic, 06130 Grasse, France
[8] Anton Pannekoek Institute, University of Amsterdam, Science Park 904, 1098XH Amsterdam, The Netherlands
[9] JILA, University of Colorado Boulder and National Institute of Standards and Technology, Boulder, CO 80309, USA
[10] Department of Physics, University of Colorado Boulder, Boulder, CO 80309, USA
[11] Department of Astronomy, University of Texas at Austin, Austin, TX 78712, USA
[12] Instituto de Astrofísica de Canarias, C/Vía Láctea S/N, E-38205 La Laguna, Spain
[13] Departamento de Astrofísica, Universidad de La Laguna, E-38206 La Laguna, Spain
[14] Departament de Física Quàntica i Astrofísica (FQA), Universitat de Barcelona (UB), c. Martí i Franquès, 1, 08028 Barcelona, Spain
[15] Institut de Ciències del Cosmos (ICCUB), Universitat de Barcelona (UB), c. Martí i Franquès, 1, 08028 Barcelona, Spain
[16] Department of Physics and Astronomy, Chalmers University of Technology, 412 96 Gothenburg, Sweden
[17] Institute for Scientific Research, Boston College, 140 Commonwealth Ave., Chestnut Hill, MA 02467, USA
[18] Université Côte d'Azur, Observatoire de la Côte d'Azur, CNRS, Lagrange, 96 Bd. de l'Observatoire, 06300 Nice, France
[19] Consejo Superior de Investigaciones Científicas (CSIC), Spain
[20] Leiden Observatory, Leiden University, PO Box 9513, 2300 RA Leiden, The Netherlands
[21] Jet Propulsion Laboratory, MS 183-900, 4800 Oak Grove Dr., California Institute of Technology, Pasadena, CA 91109, USA
[22] Space Telescope Science Institute, 3700 San Martin Dr., Baltimore, MD 21218, USA
[23] Department of Physics and Astronomy, University of North Carolina, Chapel Hill, NC 27599-3255, USA
[24] University of West Georgia, Carrollton, GA 30118, USA
[25] European Southern Observatory, Karl-Schwarzschild Strasse 2, 85748 Garching, Germany
[26] Cardiff Hub for Astrophysics Research and Technology (CHART), School of Physics and Astronomy, Cardiff University, The Parade, Cardiff CF24 3AA, UK
[27] Department of Physics, University of West Georgia, Carrollton, GA 30118, USA
[28] Jodrell Bank Centre for Astrophysics, Department of Physics & Astronomy, The University of Manchester, Oxford Rd., Manchester M13 9PL, UK

## Abstract

We report the detection of a set of new near-infrared emission features between 3.5 and 5.2 $\mu$m in JWST/NIRSpec observations of Tc 1, the planetary nebula known for displaying the cleanest and most prominent mid-infrared cosmic fullerene spectrum. These broad features share the same spatial distribution as the well-known $C_{60}$ and $C_{70}$ mid-infrared emission bands, peaking in an asymmetric ring approximately 5″–6″ from the central star. Through comparison with new anharmonic quantum chemical calculations, we demonstrate that these features arise from $C_{60}$ combination bands, marking their first detection in an astrophysical environment. The total energy radiated in the combination bands amounts to ~17% of the total energy emitted from all $C_{60}$ modes, with direct implications for fullerene cooling models. These near-infrared combination bands offer a promising new window for identifying and studying the molecular astrophysics of $C_{60}$ in sources where mid-infrared spectra are more complex.

## 1. Introduction

The buckminsterfullerene $C_{60}$ was first detected in space through its four fundamental mid-infrared emission bands at 7.0, 8.5, 17.4, and 18.9 $\mu$m in Spitzer-IRS observations of the planetary nebula (PN) Tc 1 (J. Cami et al. 2010). Tc 1 remains a premier target for detailed fullerene studies: its spectrum shows the most prominent and cleanest $C_{60}$ bands known, with additional bands attributed to its chemical cousin $C_{70}$, and furthermore with minimal blending from polycyclic aromatic hydrocarbon (PAH) emission (J. Cami et al. 2010; J. Bernard-Salas et al. 2012), making it the ideal "laboratory" for studying the molecular astrophysics of cosmic fullerenes. Unlike PAHs, for which specific molecules cannot be distinguished between species through their infrared emission, $C_{60}$ and $C_{70}$ have unique spectroscopic fingerprints that make them the only known large molecules in space amenable to detailed species-specific study. How $C_{60}$ forms, survives, and emits in harsh astrophysical environments—and precisely why it emits the way it does—remain open and fundamental questions in molecular astrophysics.

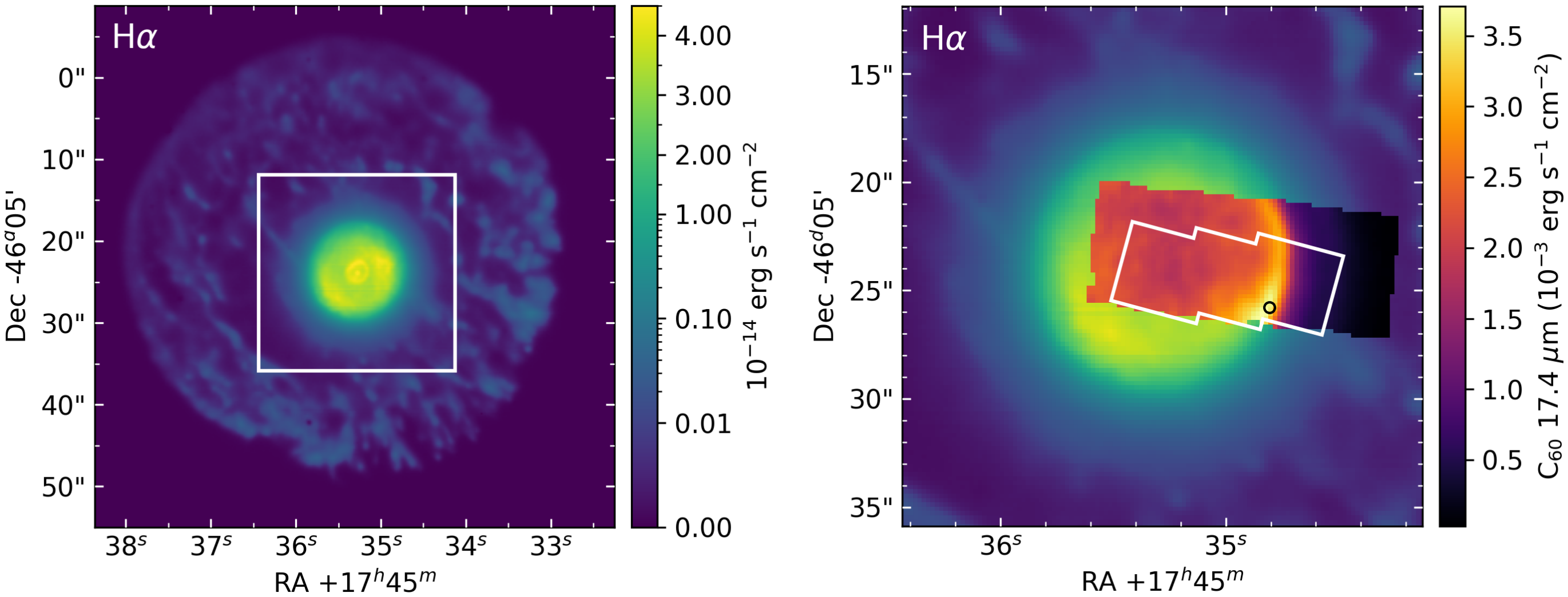

**Figure 1.** Hα maps of Tc 1 from VLT/MUSE observations (J. R. Walsh et al. 2026) showing the bright central nebula and surrounding halo. The right panel shows a close-up of the central region indicated by the white box in the left panel, with the $C_{60}$ 17.4 $\mu$m surface brightness map from the MIRI observations (M. M. Giese et al. 2026, in preparation) overlaid, revealing the fullerene emission ring. The white box in the right-hand panel indicates the footprint of the NIRSpec data discussed in this Letter. The black circle corresponds to the location of the spectrum shown in Figure 2.

As part of a comprehensive JWST observational program to study Tc 1 in greater detail, we have obtained near- and mid-infrared spectroscopy as well as imaging (see Section 2). A full analysis of this dataset will be presented in a series of forthcoming papers. Here, we report the serendipitous detection of a rich set of previously unknown emission features between 3.5 and 5.2 $\mu$m (Section 3). We characterize their spatial distribution (Section 3) and demonstrate their identification as $C_{60}$ combination bands through comparison with new anharmonic quantum chemical calculations (Section 4).

## 2. Observations and Data Reduction

Tc 1 is a young, low-excitation PN at a distance of $3530^{+380}_{-329}$ pc (N. Chornay & N. A. Walton 2021). The main ionized nebula has a diameter of $\sim$12″ and is roughly spherical, surrounded by a much fainter halo extending to $\sim$50″ (R. L. M. Corradi et al. 2003; R. Williams et al. 2008). A recent Hα map (obtained with the Very Large Telescope (VLT) MUSE; see Figure 1) clearly shows both components (J. R. Walsh et al. 2026). Spitzer-IRS and ground-based Gemini/T-ReCS observations already revealed that the $C_{60}$ emission peaks in a ring at approximately 5″–6″ from the central star (J. Bernard-Salas et al. 2012; J. Cami et al. 2018).

We observed Tc 1 as part of JWST Cycle 3 program GO-4706 (PI: J. Cami), obtaining data with both the Mid-Infrared Instrument (MIRI) using the Medium-Resolution Spectroscopy (MRS) mode (M. Wells et al. 2015; I. Argyriou et al. 2023) and the Near Infrared Spectrograph (NIRSpec; P. Jakobsen et al. 2022) using the Integral Field Unit (IFU; M. F. Closs et al. 2008). Both MIRI and NIRSpec footprints were designed to cover a region that includes the central star and a portion of the $C_{60}$ ring, and extends into the surrounding halo (Figure 1); this required a 3 × 1 mosaic for both instruments.

The NIRSpec observations were taken on 2025 April 29–30 and provide a full spectrum at each pixel covering the 0.97–5.3 $\mu$m wavelength range at a spatial scale of 0.″1 pixel$^{-1}$. The spectral range is covered by three segments (G140H-F100LP, G235h-F170LP, and G395H-F290LP); in this Letter, we focus exclusively on the G395H-F290LP segment (hereafter called "Segment 3"), which covers the 2.87–5.27 $\mu$m range. The remaining segments will be discussed in forthcoming papers. We reduced the data using the standard JWST pipeline version 1.20.2 (H. Bushouse et al. 2025) with the Calibration Reference Data System jwst_1464.pmap context file and default parameters.

## 3. Results

Figure 2 shows a typical spectrum of a pixel within the fullerene ring of the JWST/NIRSpec Segment 3 observations of Tc 1; all spectra in the ring have similar characteristics. In addition to several narrow atomic emission lines, the spectra reveal prominent, much broader emission features, not previously reported, at 3.47, 3.76, 4.3, 4.41, 4.59, 4.73, 4.82, 5.06, and 5.19 $\mu$m. We note that there is also a weak emission feature at 3.36 $\mu$m, but we excluded that feature from the further analysis and discussion due to the low signal-to-noise ratio in this wavelength region. Additionally, there is an emission feature at 3.3 $\mu$m originating from PAH emission; the weak PAH emission in this source is described in a separate paper along with a discussion on its implications on fullerene formation (S. Van Schuylenbergh et al. 2026).

To characterize these features, we decomposed them into Gaussian or Lorentzian components—whichever better reproduced each band—using the fitting software package LMFIT (M. Newville et al. 2025). We masked unresolved atomic lines, and weighted each data point by $1/\sigma_i^2$ where $\sigma_i$ is the uncertainty on the $i$th data point. We included a linear baseline underneath each component to represent the continuum. The best-fit parameters are listed in Table 1, and the fits are shown in Figure 2. Most features can be reproduced well by a single component; the strong 4.3 $\mu$m feature requires three Lorentzians, and the three components together are

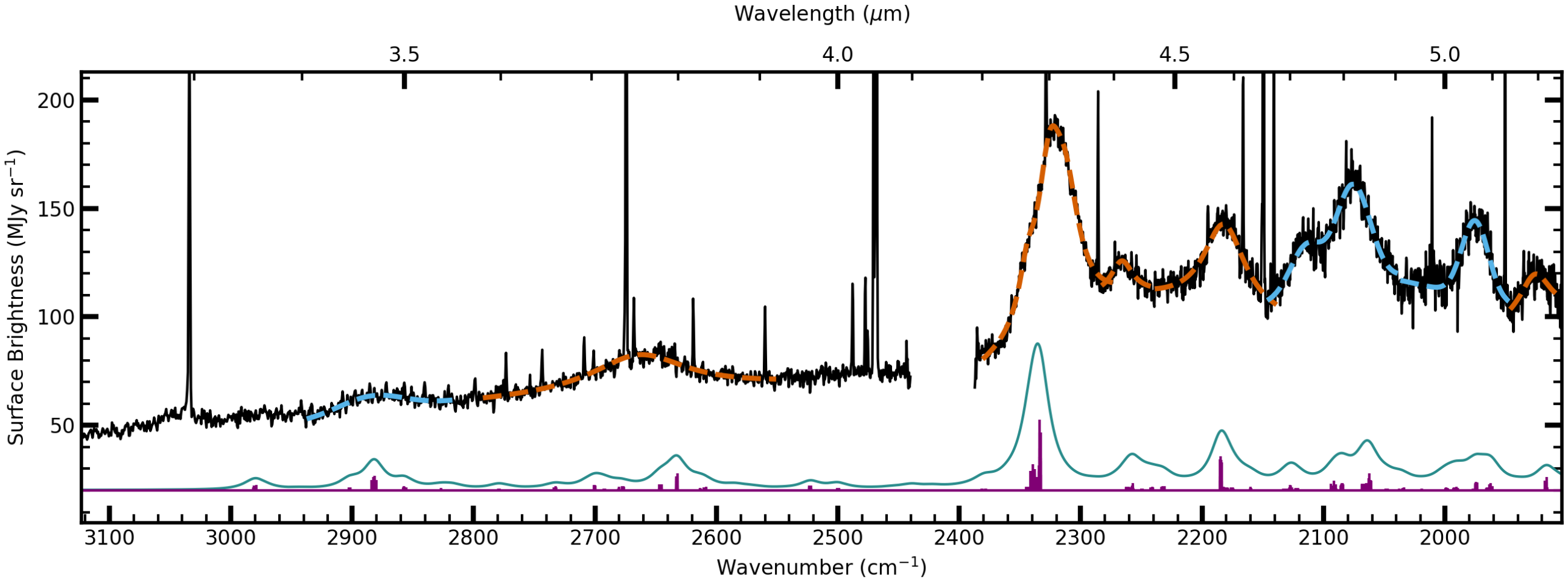

**Figure 2.** JWST/NIRSpec Segment 3 spectrum of Tc 1 (black) corresponding to a pixel within the fullerene ring; its location is denoted by a black circle in Figure 1. Gaussian (light blue, dashed) and Lorentzian (orange, dashed) fits are shown on top of the spectrum. Anharmonic calculations of $C_{60}$ combination bands are shown scaled to an arbitrary absolute intensity (teal, solid), with the stick spectrum (purple) convolved with Lorentzian profiles with line width 20 $cm^{-1}$ and shifted by $-45$ $cm^{-1}$ to match the observations.

**Table 1**
Best-fit Parameters for the Fullerene Combination Modes

| Feature | $\lambda_c$ ($\mu$m) | Width ($\mu$m) | $\tilde{\nu}_c$ ($cm^{-1}$) | Width ($cm^{-1}$) | Type |
|---|---|---|---|---|---|
| 3.47 | 3.47 | 0.03 | 2884.0 ± 2.0 | 27.0 ± 4.0 | G |
| 3.76 | 3.76 | 0.07 | 2661.0 ± 1.0 | 49.0 ± 3.0 | L |
| 4.3 | 4.27 | 0.01 | 2344.2 ± 0.8 | 8.0 ± 2.0 | L |
| | 4.30 | 0.02 | 2326.0 ± 1.0 | 13.0 ± 3.0 | L |
| | 4.32 | 0.03 | 2313.0 ± 2.0 | 14.0 ± 2.0 | L |
| 4.41 | 4.41 | 0.02 | 2266.0 ± 0.7 | 11.0 ± 2.0 | L |
| 4.59 | 4.59 | 0.03 | 2181.5 ± 0.5 | 15.0 ± 1.0 | L |
| 4.73 | 4.73 | 0.03 | 2116.1 ± 0.8 | 13.0 ± 1.0 | G |
| 4.82 | 4.82 | 0.03 | 2075.9 ± 0.4 | 14.0 ± 0.6 | G |
| 5.06 | 5.06 | 0.03 | 1975.0 ± 0.4 | 10.7 ± 0.4 | G |
| 5.19 | 5.19 | 0.04 | 1925.0 ± 1.0 | 15.0 ± 4.0 | L |

**Notes.**
[a] $\lambda_c$ and $\tilde{\nu}_c$ denote the central position in wavelength and wavenumbers, respectively, of the feature.
[b] G and L indicate a Gaussian profile and Lorentzian profile, respectively.

hereafter referenced as simply the "4.3 $\mu$m feature." The 4.73 and 4.82 $\mu$m features are blended, but have components separated enough to warrant distinction between the two.

From the fitted Gaussian and Lorentzian components at each pixel, we determined the total surface brightness for each band by integrating the fitted components, thereby excluding any blended atomic lines and continuum contributions. Figure 3 shows the integrated surface brightness map for the strongest of the new features, the 4.3 $\mu$m band. The spatial distribution of this emission closely mirrors that of the mid-IR $C_{60}$ and $C_{70}$ bands. The MIRI data reveal the $C_{60}$ emission ring in great detail, including slight asymmetries (see the map of the 17.4 $\mu$m $C_{60}$ band in Figure 1); a detailed analysis will be presented in Giese et al. (2026, in preparation). A key result from that analysis is that *all* mid-IR $C_{60}$ and $C_{70}$ emission bands in the MIRI-MRS data show the same spatial distribution: the emission is found throughout the nebula but peaks in this slightly asymmetric ring, near the outer edge of the bright core. This distribution is unique to the fullerenes: maps of recombination and fine-structure lines as well as dust continuum and dust features show very different spatial distributions, with no evidence for a ringlike structure at the same location (J. Cami et al. 2026, in preparation; M. M. Giese et al. 2026, in preparation; J. R. Walsh et al. 2026). Surface brightness maps of all newly detected near-IR bands are shown in the Appendix; all bands show the same morphology as described above for the 4.3 $\mu$m map.

## 4. Discussion

Since the spatial distribution of the near-IR bands closely follows the characteristic distribution of the fullerenes in Tc 1, the carrier of the near-IR bands must most likely be $C_{60}$, $C_{70}$, or a very closely related species. In fact, the wavelength range 3.5–5.2 $\mu$m corresponds to the wavelength range where combination bands of $C_{60}$ are expected. Unlike fundamental modes, combination bands arise from simultaneous excitation of two or more vibrational modes and cannot be accurately predicted from harmonic calculations alone—their frequencies and intensities require a full anharmonic treatment.

We therefore performed anharmonic quantum chemical calculations at the B3LYP/6-31G level of theory (W. J. Hehre et al. 1972; A. D. Becke 1993) using second-order vibrational perturbation theory (VPT2; I. Mills 1972), which allowed us to predict the positions and relative intensities of $C_{60}$ combination bands. The calculations were performed using Gaussian 16 (M. J. Frisch et al. 2016). The use of the 6-31G basis set results in modest errors in the absolute frequencies of the computed fundamental modes (experimental wavelengths at 7.0, 8.5, 17.4, and 18.9 $\mu$m and theoretical wavelengths of 6.93, 8.43, 17.3, 18.2 $\mu$m), which propagate to the combination bands; however, the overall structure and shape of the computed spectrum are not affected. An anharmonic calculation using a larger 6-31+G(d) basis set that includes polarization functions was attempted, but the computational cost rendered it inaccessible. Comparison of the harmonic frequencies computed with the 6-31G and 6-31+G(d) basis sets shows an average difference of $\sim$12 $cm^{-1}$ for all 174 modes. While the larger basis set would provide slightly more absolute accuracy, the interpretation of the observations and conclusions would not change.

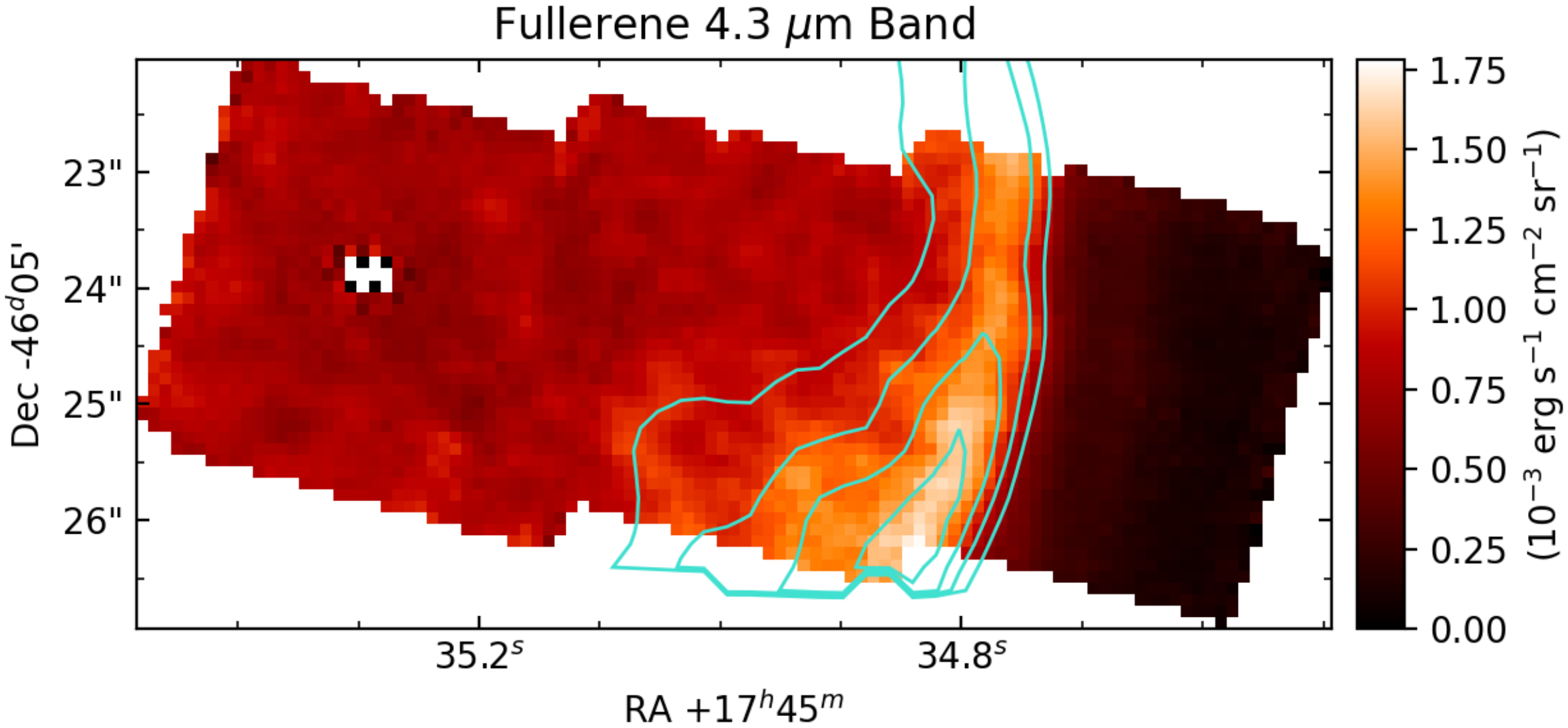

**Figure 3.** Integrated surface brightness map of the 4.3 $\mu$m band. The white patch near the central star indicates pixels affected by diffraction and fringing artifacts. Blue contours show the $C_{60}$ 17.4 $\mu$m emission from Giese et al. (2026, in preparation), demonstrating that the near-IR and mid-IR fullerene emission trace the same spatial distribution.

High-frequency modes are exclusively emitted when the molecule is highly excited and the interaction with spectator modes then introduces a redshift of the band. For PAHs, this redshift amounts to ∼20–30 cm$^{-1}$, depending on the internal energy (C. J. Mackie et al. 2022). The shifts expected for $C_{60}$ are not known but are expected to be of similar magnitude. Here, we shift the bands by 45 cm$^{-1}$ to the red to provide a visually acceptable match with the observed peak positions, where we note that the use of a low-level basis set may also introduce some uncertainty in the calculated peak positions. The resulting computed spectrum shows a remarkable correspondence with the observed features (Figure 2), and this constitutes the first comparison of anharmonic $C_{60}$ combination band calculations with observational data.

Each observed emission band comprises multiple transitions—for example, the 4.3 $\mu$m feature arises from five adjacent combination bands—and the high density of states leads to numerous vibrational resonance interactions that perturb each transition. All 174 normal modes are represented in the near-IR combination bands, with some of the most intense transitions involving the IR-active fundamental modes. The identification as $C_{60}$ combination bands is further supported by the presence of a combination band at 6.5 $\mu$m already detected in mid-IR observations of Tc 1 and other fullerene-containing objects (G. C. Sloan et al. 2014) and identified as a combination band (A. C. Brieva et al. 2016). The assignment is further supported through previous laboratory studies of $C_{60}$ films (B. Chase et al. 1992; K.-A. Wang et al. 1993; T. Hara et al. 1997). While we do not expect an exact match between our observations of gas-phase $C_{60}$ and these thin-film laboratory studies, the similar presence of features in the 1900–2400 cm$^{-1}$ adds confidence to our identification. Further detailed laboratory studies of the $C_{60}$ combination bands will be required for precise band assignments.

Given that $C_{70}$ is also detected in Tc 1 through its mid-IR emission bands, its combination bands may well contribute to the near-IR emission in this wavelength range as well. To our knowledge, no anharmonic calculations—or even laboratory studies in this wavelength range—of $C_{70}$ have been published in the literature, making it difficult to determine exact contributions. Further studies are necessary to make any assignments for $C_{70}$ contributions. Additionally, since only weak PAH emission features associated with deuterated PAHs are known at these wavelengths, the near-IR combination bands offer a promising complementary diagnostic for identifying $C_{60}$ in sources where mid-infrared spectra are more complex.

The feature at 3.47 $\mu$m (2884 cm$^{-1}$) falls near the CH stretching region, where the protonated species $C_{60}H^+$ has a band near 3.53 $\mu$m (2829 cm$^{-1}$) (L. Finazzi et al. 2024). However, the 55 cm$^{-1}$ difference between the experimental and observed peak positions rules out an assignment to $C_{60}H^+$. Furthermore, the spatial distribution of the 3.47 $\mu$m feature follows that of the other combination bands, and its relative strength is in good agreement with the calculated value, strengthening the assignment of this band to the $C_{60}$ combination bands. As a corollary, this implies that there is little contribution from $C_{60}H^+$. This is also expected on physical grounds since the reaction between $C_{60}$ and protons mainly leads to charge exchange, keeping the $C_{60}H^+$ abundance negligible.

From the combined near- and mid-infrared observational data, we find that the total energy radiated in the combination bands (including the 6.5 $\mu$m band) amounts to ∼17% of the total energy emitted in the $C_{60}$ modes, with direct implications for fullerene cooling cascade models. This nonnegligible fraction suggests that combination band emission should be accounted for in models of $C_{60}$ cooling in astrophysical environments. The combination band emission in the 4.5 $\mu$m range could furthermore explain why fullerene-containing sources in the Large Magellanic Cloud contain similar K–[3.6] and [4.5]–[8] colors (G. C. Sloan et al. 2014). The relative intensities of the combination bands compared to the fundamental modes also provide new observational constraints on the average internal energy of the $C_{60}$ molecules, and thus on the excitation mechanism responsible for their emission.

## 5. Conclusion

We report the first detection of $C_{60}$ combination bands in an astrophysical environment, identified in JWST/NIRSpec observations of the PN Tc 1. The spatial distribution of these near-IR features matches that of the mid-IR $C_{60}$ and $C_{70}$ emission bands, and their identification is confirmed by new anharmonic quantum chemical calculations. The combination bands carry ∼17% of the total energy emitted in all $C_{60}$ modes, suggesting that they should be accounted for in models of $C_{60}$ cooling in astrophysical environments. Since only weak PAH emission features associated with deuterated PAHs are known at these wavelengths, the near-IR combination band region offers a promising diagnostic window for identifying $C_{60}$ in sources where features from other molecular and dust features interfere with distinguishing the mid-infrared bands. Laboratory measurements of $C_{60}$ combination band positions would be valuable for refining the identifications presented here.

## Acknowledgments

This work is based on observations made with the NASA/ESA/CSA James Webb Space Telescope. All of the data presented in this Letter were obtained from the Mikulski Archive for Space Telescopes (MAST) at the Space Telescope Science Institute. The data of this specific observing program can be accessed via doi: 10.17909/5qz6-ay45. This research is based upon work from COST Action NanoSpace, CA21126, supported by COST (European Cooperation in Science and Technology). M.G., J.C., S.V.S., E.P., C.B., and D.V.D.P. acknowledge support from the University of Western Ontario, the Canadian Space Agency (CSA) [24JWGO3A13], and the Natural Sciences and Engineering Research Council of Canada. V.J.E. thanks Chapman University for support. D.A.G.H. and A.M. acknowledge the support from the State Research Agency (AEI) of the Ministry of Science, Innovation and Universities (MICIU) of the Government of Spain, and the European Regional Development Fund (ERDF), under grant PID2023–147325NB–100/AEI/10.13039/501100011033. Astrochemistry in the Netherlands is supported by the NWO Dutch Astrochemistry Network (grant No. ASTRO.JWST.001). R.S.'s contribution to the research described here was carried out at the Jet Propulsion Laboratory, California Institute of Technology, under a contract with NASA (80NM0018D0004). Support for participation in JWST-GO-04706 by G.C.S., H.L.D, and R.S. was provided through grants from the STScI under NASA contract NAS5-03127. P.B.C. is supported by NIST. K.J. acknowledges the support of the Swedish National Space Agency.

*Software*: ASTROPY (Astropy Collaboration et al. 2013, 2018), LMFIT (M. Newville et al. 2025).

## Appendix
## Fullerene Combination Band Spatial Maps

Figure 4 shows the surface brightness spatial distribution of all detected $C_{60}$ combination bands in the near-IR. All maps show a similar distribution in which the emission peaks in an asymmetric ring near the edge of the bright core. We note that some fits for the weaker bands are severely affected by noise—most evident in the 3.47, 4.41, and 5.19 $\mu$m maps.

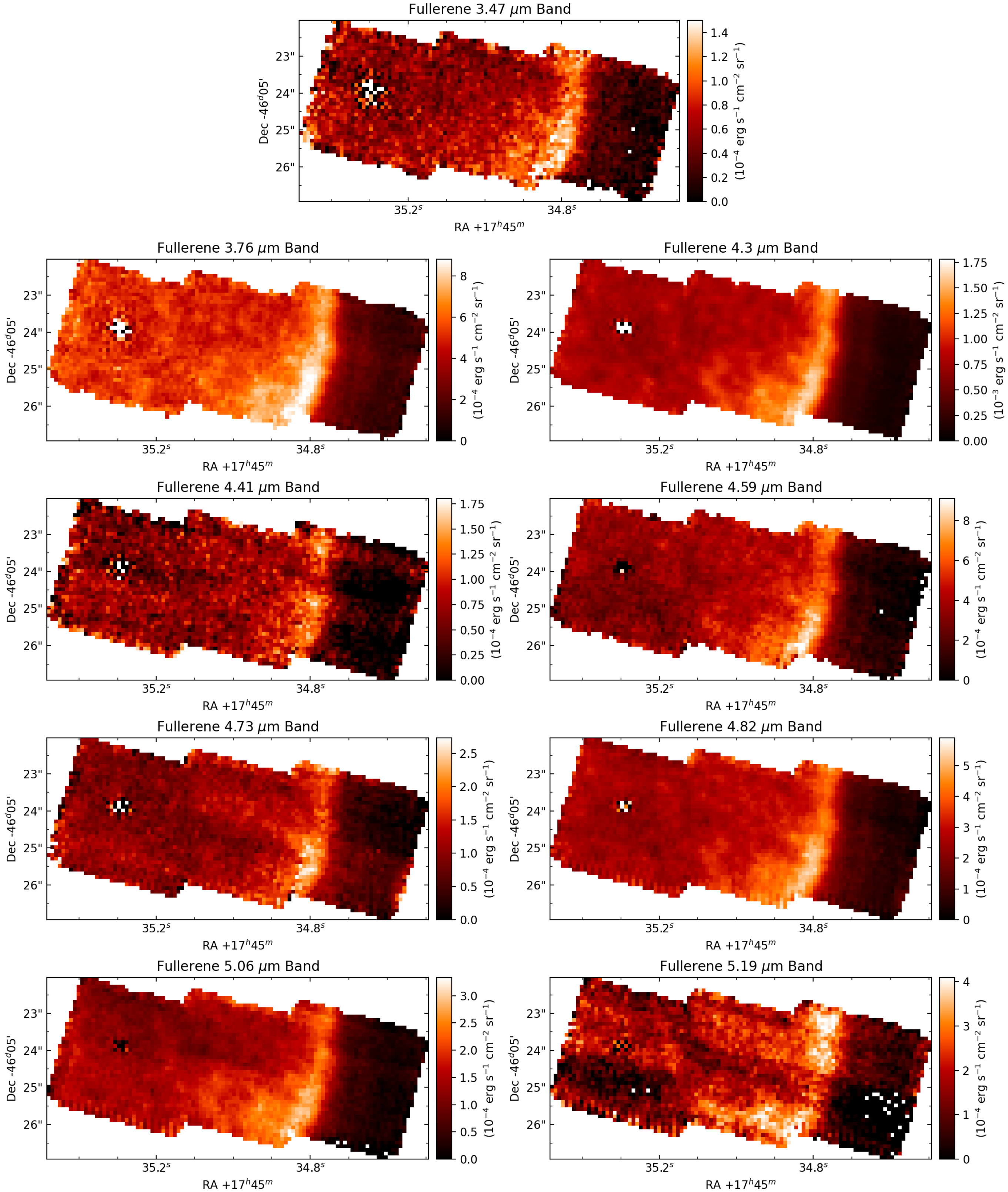

**Figure 4.** Integrated surface brightness maps of each of the fullerene features. The white/black patch near the central star in each map indicates pixels affected by diffraction and fringing artifacts.

## ORCID iDs

Morgan M. Giese https://orcid.org/0009-0007-3950-335X
Vincent J. Esposito https://orcid.org/0000-0001-6035-3869
Simon Van Schuylenbergh https://orcid.org/0009-0004-9270-0019
Jan Cami https://orcid.org/0000-0002-2666-9234
Els Peeters https://orcid.org/0000-0002-2541-1602
Charmi Bhatt https://orcid.org/0009-0000-0191-6756
Dries Van De Putte https://orcid.org/0000-0002-5895-8268
A. G. G. M. Tielens https://orcid.org/0000-0003-0306-0028
Michael J. Barlow https://orcid.org/0000-0002-3875-1171
Jeronimo Bernard-Salas https://orcid.org/0000-0002-8452-8675
Alessandra Candian https://orcid.org/0000-0002-5431-4449
Bryan Changala https://orcid.org/0000-0003-0304-9814
Nick L. J. Cox https://orcid.org/0000-0002-7926-4492
Harriet L. Dinerstein https://orcid.org/0000-0002-4017-5572
D. A. García-Hernández https://orcid.org/0000-0002-1693-2721
Marco A. Gómez-Muñoz https://orcid.org/0000-0002-3938-4211
Kay Justtanont https://orcid.org/0000-0003-1689-9201
Kathleen E. Kraemer https://orcid.org/0000-0002-2626-7155
Eric Lagadec https://orcid.org/0000-0002-1335-5623
Arturo Manchado https://orcid.org/0000-0002-3011-686X
Ana Monreal Ibero https://orcid.org/0000-0002-6455-2491
Raghvendra Sahai https://orcid.org/0000-0002-6858-5063
Ameek Sidhu https://orcid.org/0000-0003-3771-4990
G. C. Sloan https://orcid.org/0000-0003-4520-1044
N. C. Sterling https://orcid.org/0000-0002-9604-1434
Jeremy R. Walsh https://orcid.org/0000-0002-8008-910X
Roger Wesson https://orcid.org/0000-0002-4000-4394
Albert Zijlstra https://orcid.org/0000-0002-3171-5469